\documentclass{article}
\usepackage{arxiv}
\usepackage[authoryear]{natbib}
\usepackage{url}
\usepackage{hyperref}

\usepackage{placeins}

\usepackage{graphicx}%
\usepackage{multirow}%
\usepackage{amsmath,amssymb,amsfonts}%
\usepackage{amsthm}%
\usepackage{mathrsfs}%
\usepackage[title]{appendix}%
\usepackage{xcolor}%
\usepackage{textcomp}%
\usepackage{manyfoot}%
\usepackage{booktabs}%
\usepackage{algorithm}%
\usepackage{algorithmicx}%
\usepackage{algpseudocode}%
\usepackage{listings}%
\usepackage{svg}%
\usepackage{xspace}

\theoremstyle{definition}

\newtheorem{definition}{Definition}

\newtheorem{remark}{Remark}

\newcommand{\N}{\mathbb{N}}
\newcommand{\Nset}[1]{\mathbb{N}_{#1}}
\newcommand{\R}{\mathbb{R}}
\newcommand{\Rnn}{\mathbb{R}_{\ge 0}}
\newcommand{\abs}[1]{\lvert #1 \rvert}
\newcommand{\flowfunc}[1]{\mathfrak{F}(#1)}
\newcommand{\flowfuncnull}{\mathfrak{F}}
\newcommand{\pardim}[1]{\mathfrak{p}(#1)}
\newcommand{\stadim}[1]{\mathfrak{s}(#1)}
\newcommand{\betavec}{\vec{\beta}}
\newcommand{\xvec}{\vec{x}}
\newcommand{\fr}{\ensuremath{\mathcal F}}
\newcommand{\tauf}{\tau^{\mbox{\tiny from}}}
\newcommand{\taua}{\tau^{\mbox{\tiny arg}}}
\newcommand{\define}[1]{\emph{{#1}}}
\newcommand{\vv}{{\emph{vice versa}\xspace}}

\begin{document}

\title{Toward a comprehensive system for constructing compartmental epidemic models}

\author{Darren Flynn-Primrose \\
  Department of Mathematics \& Statistics, McMaster University \\
  \texttt{flynnprd@mcmaster.ca} \\
  \And Steven C. Walker \\
  Department of Mathematics \& Statistics, McMaster University \\
  \And Michael Li \\
  Public Health Risk Science Division, National Microbiology Laboratory, Public Health Agency of Canada \\
  \And Benjamin M. Bolker \\
  Departments of Biology and Mathematics \& Statistics, McMaster University \\
  \And David J.\,D. Earn \\
  Department of Mathematics \& Statistics, McMaster University \\
  \And Jonathan Dushoff \\
  Department of Biology, McMaster University
}

\maketitle
    
\begin{abstract}
  Compartmental models are valuable tools for investigating infectious diseases. Researchers building such models typically begin with a simple structure where compartments correspond to individuals with different epidemiological statuses, e.g., the classic SIR model which splits the population into susceptible, infected, and recovered compartments. However,
  as more information about a specific pathogen is discovered, or as a means to
  investigate the effects of heterogeneities, it becomes useful to stratify models further --- for example by age, geographic location, or pathogen strain. The operation of constructing stratified compartmental models from a pair of simpler models resembles the Cartesian product used in graph theory, but several key differences complicate matters. In this article we give explicit mathematical definitions for several so-called ``model products'' and provide examples where each is suitable. We also provide examples of model stratification where no existing model product will generate the desired result. 
\end{abstract}


\section{Introduction}\label{intro}
The COVID-19 pandemic has reemphasized the importance of compartmental models \citep{abou2020compartmental, massonis2021structural, adam2020special, currie2020simulation, lofgren2014mathematical, mcbryde2020role, enserink2020covid} and has resulted in a flood of new compartmental models (e.g., \cite{friston2020dynamic, fields2021age, chang2022stochastic, lavielle2020extension, balabdaoui2020age, leontitsis2021seahir}). 
This abundance of new model variants is to be expected given the number of public health modelers seeking to integrate the current scientific understanding of emerging infectious diseases in a way that will have policy impact. Modelers must be able to build models rapidly to explore scenarios and generate high quality forecasts; public health recommendations have the biggest impact if they can be acted on promptly. However, the speed at which modelers can develop new models typically trades off with model quality. It is therefore useful to develop tools that make it easier for modelers to build high-quality models more quickly.

One way to address this speed-quality trade-off is to build infectious disease models incrementally. Information is scarce early in an epidemic, and so early models should be simple to reflect ignorance. As epidemics progress, more information is gathered and policy choices require fast input from scientists. Public health modelers will then need to quickly add complexity to their models if they are to be relevant to policy. 

Proceeding in this way can eventually result in extremely complex models, much of whose complexity is no longer relevant. Complexity makes it more difficult to add new features to the model when they become necessary. Therefore, modelers need tools that make it easier to flexibly add and remove model features. 

\cite{savageau1988introduction} and \cite{voit1988recasting, voit1990s} made an early attempt to create such a toolbox by recasting the underlying differential equations of a model into a canonical form they call an ``S-model". Unfortunately this effort focused on the model's differential equations rather than its graphical structure, thus making it unsuitable for less mathematically inclined modelers. It does not seem to have been widely adopted.

\cite{friston2020dynamic} describe how the state space of a complex epidemiological model can be constructed from the product of different latent state dimensions (their Figure 1: infection status, clinical status, testing status, and location), but the definition of which compartments are connected, and the rates of flow between them, is left up to the modeller.

A recent and promising effort to formalize the construction of compartmental models employs the language of category theory \citep{fong2018seven, Libkind2022an, libkind2021operadic, baez2022compositional, baez2017compositional}. This powerful approach addresses many of the concepts we discuss here; however, at its current stage of development it requires considerable knowledge of advanced mathematics to use effectively.  An ongoing project to implement the category theoretic approach in the Julia language can be found at \url{https://github.com/AlgebraicJulia/AlgebraicPetri.jl} \citep{algebraicjulia}. 

\cite{worden2017products} use the relatively simple language of graph theory to describe common methods of combining compartmental models. The current paper is a result of the our efforts to implement the products described by Worden and Porco in software. At a high level, we view model ``multiplication'' as a three-step operation:

\paragraph{Procedure for Model Multiplication}\label{genproc}

\begin{enumerate}
    \item Generate the vertices of the product model by combining the vertices of the factor models. This typically means taking the Cartesian product of the vertices of factor models. In many cases, we will want only a subset of the Cartesian product in the final model (e.g., some combinations are physically or biologically impossible).
    \item Generate the edges of the product model. Again, we will typically take the Cartesian product of edges in each factor model with the vertices in the other. Some transitions may be disallowed, in which case we would drop those edges from the product or set the flows across them to zero. In other cases, we may want to add edges to the product to allow state changes in multiple strata to occur simultaneously.
    \item Resolve ambiguities in how flow functions are generalized to accommodate the presence of additional strata.
    \item Set parameters for the product model
\end{enumerate}

The combination of the first two steps is just a graph-theoretic Cartesian product. Recognizing the need to adjust flow rates (the third step above) is Worden and Porco's contribution. This requirement can arise in a number of ways; for example, when stratifying a standard SIR model, we need to decide whether the susceptible population of one stratum can be infected by the infectious population of another stratum.
If the stratification is based on age groups then it is reasonable to let different strata infect one another, i.e. old people infecting young people and \vv. On the other hand, if the populations in different strata are isolated from each other it may be preferable to prohibit inter-stratum infection. Our approach follows Worden and Porco's; when computing the magnitude of a flow between compartments we separately compute the contribution from each individual stratum and then sum the resulting quantities to find the total magnitude of the flow. There are other possible approaches which we will revisit in Section \ref{aff}, but this additive approach appears to be the most common. When stated in these simple terms it is easy to imagine that we have solved the problem of combining models, but the devil is in the details. It is one thing to develop a framework for combining models and their various components but it is a much greater challenge to determine precisely which models can (and cannot) be expressed within that framework. 

Given the diversity of the audience potentially interested in this material we have attempted to strike a healthy balance between mathematical rigour and intuitive explanation. In Section \ref{dcm} we introduce some mathematical terminology that will be used in the remaining sections to rigorously define a number of distinct model products. Sections \ref{worden} and \ref{general} share similar structure, in the first subsection we focus on mathematical details and in the second subsection we provide examples which we hope will communicate our main points in a more intuitive way. Section \ref{worden} restates two products introduced by \cite{worden2017products}, Section \ref{general} describes a generalization of the products from the previous section. Section \ref{unco} has three subsections each dedicated to different complexity that can make models challenging to construct using products alone. We conclude with Section \ref{conc} where we summarize our primary results and discuss the potential for further investigation.

\section{Defining Compartmental Models}\label{dcm}

Compartmental models closely resemble directed graphs, which are graphs where the connections between nodes have a specific direction \citep{roberts2009applied}; in this analogy the directional connections correspond to flows between compartments. The relationship between directed graphs and compartmental models has been studied before, for example by \cite{walter1999compartmental} --- however, such investigations have been largely limited to the case where the magnitudes of flows between compartments are governed by linear equations. While certainly an important special case, this framework is insufficient for a disease-transmission model. In fact, a central point we hope to communicate in this paper is that, while constructing the nodes and edges of compartmental models is straightforward, making choices about how to calculate flows between compartments is a challenge with considerable nuance.

We consider a compartmental model with $n$ compartments. Flows can go from one compartment to another, from outside the system into a compartment (e.g., births), or from a compartment to the outside (e.g., deaths). There are thus up to $n(n+1)$ possible flows, although most of these will be missing in any particular model. Each flow that is present will be described by a function that may depend on the state of any of the $n$ compartments and on any of $m$ (possibly time-varying) parameters. Flows out of a compartment can be either \define{per capita} flows, in which case their value is multiplied by the current state value of that compartment, or they may be \define{absolute} flows where no further computation is required.

Historically much of the work done investigating compartmental models through a mathematical lens has been focused on the case where flow rates between compartments are restricted to linear equations (i.e., flows strictly proportional to the current state of the originating compartment).
Formally, we let $\fr(n, m)$ denote the set of allowable flow-rate functions $f:\Rnn^n\times\R^m\rightarrow \Rnn$; we will often abbreviate this as \fr. More generally, we write the dimension of a model \textit{state space} as $n=\stadim{f}$ and of the  \textit{parameter space} as $m=\pardim{f}$. It is also frequently convenient to write functions in the form $f(\vec{\beta}, \vec{x})$ where $\vec{\beta}$ is a vector in $f$'s parameter space and $\vec{x}$ is a vector in $f$'s state space. 

To move from flow-rate functions to actual flows, it is convenient to define a family of \define{filtering functions} which select specific subsets of the state space. We have two slightly different uses for these filtering functions: (1) we use them to select the ``from'' compartment of per-capita flows and (2) they will be used later to select compartments of a model product that belong to particular strata, for example selecting all compartments that correspond to a specific age group or location. We denote these families with the symbol $\Upsilon$; for example $\Upsilon(3,2)$ is the set of all simple projections from $\R^3$ to $\R^2$ (in this case, the three ways of choosing two variables from a list of three). We also include in all such families the projection onto unity to allow for the possibility of flows specified in absolute rather than per-capita terms. A filtering function being used to select a single ``from'' compartment for a per-capita flow will be denoted with $\tauf$; one being used to select all compartments from a single stratum of the model, which will typically then be used as arguments to a flow function, will be denoted with $\taua$.

We now have two families of functions. The first family is a set of flow rate functions, $\fr$, which we will use to describe either the per-capita or absolute rate of flow between compartments. The second family is a set of filtering functions, $\Upsilon$, which will be used to select the population of the ``from'' compartment for a given per-capita flow from all compartments in a model (in the case of absolute flow rates this function will map all inputs to 1). With these definitions we can now define our third and final family of functions which will be denoted by $\flowfuncnull$ and is used to describe the absolute rate of flow between compartments. Every function in $\flowfuncnull$ can be expressed as a product of one function in $\Upsilon$ and another in $\fr$. That is, if $f$ is a member of $\flowfuncnull$ then there must be a $\tauf$ in $\Upsilon$ and an $f_r$ in $\fr$ such that $f = \tauf \cdot f_r$. Since $\Upsilon$ contains functions that select the population of a single compartment as well as the projection of state space onto the number $1$, $\flowfuncnull$ will contain functions that describe the magnitude of a flow both in per-capita terms and in absolute terms. In the former case we will have $f = \tauf\cdot f_r$ where $\tauf$ corresponds to the population of the from compartment;  in the latter case we will have the same equation but $\tauf$ will reduce to the number $1$, leaving $f_r$ as the absolute flow rate between compartments. Below we give a more technical definition.

\begin{definition}[Flow Functions]\label{flowfunc}
     Given $n,m\in\N$ we define $\flowfunc{n,m}$ to be the set of all functions $f:\Rnn^n\times\R^m\rightarrow\Rnn$ such that for some $f_r\in\fr(n,m)$ and $\tauf\in\Upsilon(n, 1)$, for every $\betavec\in\R^m$ and $\xvec\in\Rnn^n$
     \begin{equation}
        f(\betavec, \xvec) = \tauf(\xvec)\cdot f_r(\betavec, \xvec)
     \end{equation}
\end{definition}

\begin{remark}[Notation, Ordered sets and Labels]
    Where sets are labeled we assume their labels correspond to the order of the set (i.e. the first element in the set is labeled “1" and the $i$\textsuperscript{th} element is labeled ``i"). Where sets are associated with each other we assume their labels (and/or order) correspond to the association map (i.e. if $A$ and $B$ are associated then $a_i\in A$ maps to $b_i\in B$ and vice versa).  
\end{remark}

A closely related concept to compartmental models is that of directed graphs (aka \textbf{digraphs}). Intuitively, digraphs are just graphs where the connections between vertices have a specific direction associated with them. A more technical definition is that a digraph consists of two sets: a set of vertices $V$ and a set of ordered pairs of vertices $E$ that defines directed connections between vertices. As we will see below, compartmental models are digraphs with an additional set of functions that regulate the magnitude of the flow across edges. Later when we discuss model products it will be helpful to recognize that the digraph underlying a product model is the Cartesian product of the digraphs underlying the factor models. First, though, we define compartmental models in the language developed thus far.

\begin{definition}[Compartmental Models]\label{compmodel}
    Let $D_0=(V,E)$ be a labeled, finite digraph and let $n=\abs{V}$. Suppose there is a set $F\subset\flowfunc{n}$ associated with $E$, Then we call $D=(D_0, F)$ a \textbf{compartmental model}. We call $\Rnn^n$ the \textbf{state space of $D$} and if $m=\sum\limits_{f\in F}\pardim{f}$, then we call $\R^m$ the \textbf{parameter space of $D$}.
\end{definition}

\subsection{Parameter space of factor and product models}
In Definition \ref{compmodel} and in much of the rest of this paper we assume the product model parameters are entirely independent of each other. So if one factor model has $k_1$ compartments and $l_1$ parameters and the other factor model has $k_2$ compartments and $l_2$ parameters then every the product model will have a minimum of $k_1l_2+k_2l_1$ parameters (when no interaction between strata is allowed) and a maximum of $k_1^2l_2+k_2^2l_1$ parameters (when all strata are allowed to interact with each other).  This approach has the advantage of preserving generality. 

In practice, parameters in the product model are often  related to parameters in the original model factors in simple mechanistic ways. 
However, there is an enormous range of possible relationships between the parameters of the factor models and the parameters of their product.
Some parameters, such as those describing intrinsic properties of a pathogen, may be constant across all strata of a product model. Others such as recovery time my be constant with respect to some dimensions of stratification (e.g. location) but variable with respect to others (e.g. age). In other cases the relationships may depend on the degree of available data --- we may know that recovery time varies with age in reality, but choose to treat is as constant for modeling purposes. It does not appear possible, with just the information present in the factor models, to deduce the desired relationship between factor model and product model parameters. Thus we choose for now to default to the most general possible case and trust that where more convenient relationships between parameters exist modelers will construct appropriate mappings for themselves. That said, we will discuss a few common scenarios here for the purpose of illustration. 

A parameter may be constant across multiple strata of a product model. This is common for parameters that describe intrinsic properties of the pathogen being modeled, where the strata represent variation among hosts. A related case occurs when the value of a parameter at each stratum in the product model is related to the factor model version of the parameter by a simple scalar. For example a pathogen may have an average recovery time across the entire population and the specific recovery time for different age groups could be specified as a percentage of the overall average. In both these cases if we let $\alpha\in\R$ denote the parameter value in the factor model and $\betavec\in\R^k$ denote the values of the derived parameters at $k$ different stratum in the product model then we can write $\betavec = \alpha\vec{w}$ where $\vec{w}\in\R^k$ is a vector of weights.

Multiple parameters in a factor model may be related to each other, as when a single compartment has multiple flows emanating from it. Consider the case of a person who has been exposed to a pathogen. Some models will allow them several possible fates: for example they could go on to be asymptomatic or have mild or severe symptoms. In that case the factor model in question will have three parameters ($\alpha_1, \alpha_2, \alpha_3\in (0,1)$) related by the fact that their sum must be one ($\alpha_1+\alpha_2+\alpha_3 = 1)$. In other words, if $\alpha_1$ and $\alpha_2$ are given then $\alpha_3 = 1-\alpha_1 - \alpha_2$. When included in a product model every stratum of the new model will have three parameters derived from the original $\alpha$ values; however each stratum may have different values for those parameters. For example people in different age groups may be more or less likely to experience severe, mild, or no symptoms. Mathematically, if we say $\vec{\beta_1}, \vec{\beta_2}, \vec{\beta_3}$ denote the parameters at every stratum of the product model derived from $\alpha_1, \alpha_2, \alpha_3$ respectively, then we can write $\vec{\beta_1} = \alpha_1\vec{w_1}$, $\vec{\beta_2} = \alpha_2\vec{w_2}$, and $\vec{\beta_3} = \vec{1} - \vec{\beta_1} - \vec{\beta_2}$.

A more complex case occurs when different strata of a product model interact. For example consider a simple SI model, that is, a model with only susceptible and infected compartments. In the standard formulation the force of infection of such a model is given by $\Lambda = \frac{\beta I}{N}$, so the total number of newly infected people is $S\cdot \Lambda = \frac{\beta S I}{N}$. Suppose we now stratify this model to represent a scenario where each person lives in one of three different locations but may come in contact with people living in the other locations. Our model would then have three infected compartments ( i.e. $\vec{I} = (I_1, I_2, I_3)$) and three susceptible compartments (i.e. $\vec{S} = (S_1, S_2, S_3)$; the force of infection would be generalized to a vector with three entries, one for each location (i.e. $\vec{\Lambda} = (\lambda_1, \lambda_2, \lambda_3)$). 
In the most general case, where the force of infection does not take the standard form given above, each $\lambda_i$ would be expressed as \emph{some} function of the infected populations as well as a vector of parameters $\vec{\beta_i}$ which gives some information about how people at different locations interact with each other. Thus we would be left with $\vec{\Lambda} = (f(\vec{\beta_1}, \vec{I}), f(\vec{\beta_2}, \vec{I}), f(\vec{\beta_3}, \vec{I}))$. 
In the standard formulation the force of infection is a linear equation with respect to the population of infected compartments, so we can be more specific. The factor model parameter $\beta$ is generalized to nine new, presumed unrelated, parameters which can be written as a $3 \times 3$ matrix
\[
    B = \begin{pmatrix}
        \beta_{11} & \beta_{12} & \beta_{13} \\
        \beta_{21} & \beta_{22} & \beta_{23} \\
        \beta_{31} & \beta_{32} & \beta_{33}
    \end{pmatrix}
\]
with the end result that we can write the expression
\[
    \vec{\Lambda} = \frac{1}{N} B \vec{I}
\]
While it still preserves a degree of generality, this approach unfortunately expands the model's parameter space significantly. In practice the likelihood of a person residing in one location coming into contact with a person in a different location is not arbitrary but can reasonably be expected to vary according to the distance between the two locations. So if $D$ is a three-by-three matrix where the $d_{i,j}$ entry denotes the distance between location $i$ and location $j$ then it is possible to construct a contact matrix $C$ that assigns numerical values to the likelihood of contact between people at two locations based on the known distance between them. One way to do this would be to let every entry $c_{i,j}$ in $C$ be given by
\[
    c_{i,j} = e^{-\gamma d_{i,j}}
\]
where $\gamma \in \Rnn$ is some fixed parameter. In this way we can write
\[
    \vec{\Lambda} = \frac{\beta}{N} C \vec{I}
\]
which preserves the original meaning of the parameter $\beta$ and only introduces one new parameter ($\gamma$) instead of nine. Notice that it is possible to translate between the two approaches using the map
\[
    B = \beta C = \beta \begin{pmatrix} 
                        e^{-\gamma d_{1,1}} & e^{-\gamma d_{1,2}} & e^{-\gamma d_{1,3}} \\
                        e^{-\gamma d_{2,1}} & e^{-\gamma d_{2,2}} & e^{-\gamma d_{2,3}} \\
                        e^{-\gamma d_{3,1}} & e^{-\gamma d_{3,2}} & e^{-\gamma d_{3,3}} \\
                        \end{pmatrix}
\]

There are of course other ways to handle this kind of parameter simplification (e.g. \cite{andemay85, andemaybook, grenande85}). Most situations will allow for a parameter space mapping of this kind that relates the default parameter space generated by model products to a smaller parameter space dictated by the specific data available to the modeler. However, as we have tried to make clear in this subsection, the sheer variety of parameter space mappings that could be useful in some situation is so great that for now we will maintain generality by treating all parameters in the product model as independent.

In our general parameterization the $l_i$ parameters in the product model that come from the $i$'th factor model (of two) can be organized in a $l_i \times k_{2-i} \times k_{2-i}$ order three tensor, $B^{(i)}$. The modeler will have some set of parameters known to them which we call $\vec{\theta}$ and will be able to compose, from a library of standard relations, a mapping $g$ so that
$$
B_{hij}^{(i)} = g_{hij}^{(i)}(\vec{\theta})
$$

\section{Naive and Modified Products}\label{worden}

\cite{worden2017products} describe several distinct model products, two of which are particularly relevant to us. Here, we would like to restate these two definitions in a mathematically rigorous way. These formal definitions provide a foundation for us to define a new, generalized version of a model product in Section \ref{general} and have been valuable in our efforts to automate the process of model stratification. However, we will begin with a description of the Cartesian product of digraphs as they are a significant component of model products; treating them separately will simplify our later definitions.

\begin{definition}[Cartesian Product of Digraphs]\label{cpd}
    Let $D^a = (V^a, E^a)$ and $D^b = (V^b, E^b)$ be digraphs with vertices $V^{\bullet}$ and edges $E^{\bullet}$. Then the \textbf{Cartesian Product} of $D^a$ and $D^b$ (i.e. $D^a\times D^b$) can be written as $D=(V, E)$ where $V=V^a\times V^b$ and $E$ is the union of two sets $E^c$ and $E^d$ which are given by
    \begin{equation}
        E^c = \left\{((w,x),(y,x))\in V\times V \vert (w,y)\in E^a\right\}
    \end{equation}
    and
    \begin{equation}
        E^d = \left\{((w,x),(w,z))\in V\times V \vert (x,z)\in E^b \right\}
    \end{equation}
    respectively.
\end{definition}

It is not always the case that the digraph underlying a product model is the Cartesian product of the digraphs underlying the factor models, one example is the so-called “strong product" defined in \cite{worden2017products} which includes additional edges that are not present in the Cartesian product. Alternatively, as we shall see in Section \ref{wp}, there are some cases where the digraph underlying the product model is a proper subset of the Cartesian product of the factor models. This is particularly the case when combining models with different pathogen strains while disallowing the possibility of being infected by multiple strains simultaneously. However we believe the practical applications of these non-Cartesian model products are relatively limited; in this article we focus on model products where the related digraph product is Cartesian. \cite{worden2017products} discuss two Cartesian-like products; since the underlying digraphs of these products are identical, the only difference between them is the set of flow functions in the product model (with respect to the \hyperref[genproc]{procedure for model
  multiplication} they agree on the first two steps and only differ on the third). Notice that in Definition~\ref{cpd} the edge set of the product digraph is the union of two other sets $E^c$ and $E^d$. The set of flow functions of a model product $F$ will be formed by a similar union of sets $F^c$ and $F^d$, where $F^c$ contains the flow functions related to edges in $E^c$ and $F^d$ contains those related to $F^d$.

\begin{figure}
    \centering
    \includegraphics[width=\textwidth]{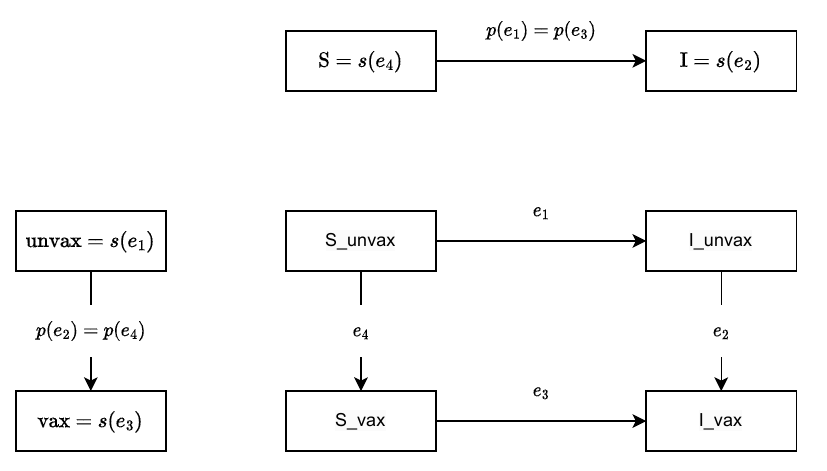}
    \caption{A Susceptible-Infected model combined with a Unvaccinated-Vaccinated model. Every edge in the product model maps to an edge in one of the factor models (via the mapping $p$) and a vertex in the other factor model (via the mapping ``s''.)
    }
    \label{fig:spexplanation}
\end{figure}

\begin{definition}[Naive and Modified Products]\label{products}
 Suppose we have two compartmental models $D^a = (V^a, E^a, F^a)$ and $D^b = (V^b, E^b, F^b)$. Let $V$, $E^c$, $E^d$, and $E$ be as in Definition \ref{cpd}. Since $E$ is the union of $E^c$ and $E^d$ every edge $e\in E$ is associated with an edge in one factor model and a vertex in the other; we use $p(e)$ to denote the associated factor model edge and $s(e)$ to denote the associated factor model vertex. For example if $e\in E^c$, then $p(e) = (w,y)\in E^a$ and $s(e) = x \in V^b$ (Diagram \ref{fig:spexplanation}). By Definition \ref{compmodel} for every $e\in E^a$ there are functions $f^a_{e}\in F^a$ and $f_{r,e}\in \fr(\abs{V^a})$, as well as a filtering function $\tauf_{e}\in\Upsilon(\abs{V^a}, 1)$ such that
 \begin{equation}
        f^a_{e}(\betavec, \xvec) = \tauf_e(\xvec)\cdot f_{r,e}(\betavec, \xvec).
 \end{equation}
 Let $\taua_e$ be the filtering function from $V$ to $V^a\times s(e)$. For every $e\in E^c$ and $\vec{x}\in\Rnn^{\abs{V}}$ we define $f^1_e:\R^{\pardim{f^a_{p(e)}}}\times\Rnn^{\abs{V}}\rightarrow\Rnn$ by
 \begin{equation}
        f^1_e(\betavec, \xvec) = \tauf_{p(e)}(\taua_e(\xvec))\cdot f_{r,p(e)}(\betavec, \taua_e(\xvec))
 \end{equation}
 and we define $F^{1,c} = \bigcup\limits_{e\in E^c}f^1_e$. A symmetric definition can be given for $e\in E^d$ yielding $F^{1,d} = \bigcup\limits_{e\in E^d}f^1_e$. We define $F^1 = F^{1,c}\cup F^{1,d}$ and we say $D^1 = (V, E, F^1)$. We define
 \begin{equation}
        D^a\boxminus D^b = D^1
 \end{equation} We call $\boxminus$ \textbf{the naive product}.

 Alternatively, we can for every $v^b\in V^b$ let $\taua_{v^b}$ be the filtering function from $V$ to $V^a\times v^b$. For every $e\in E^c$ and $\vec{x}\in\Rnn^{\abs{V}}$ we define $f^2_e:\R^{\pardim{f^a_{p(e)}}\times \abs{V^b}}\times\Rnn^{\abs{V}}\rightarrow\Rnn$ by
 \begin{equation}\label{modified_eq}
    f_e(\vec{\beta}, \vec{x}) = \tauf_{p(e)}(\taua_{s(e)}(\xvec))\sum\limits_{v^b\in V^b} f_{r,p(e)}(\betavec_{v^b},\taua_{v^b}(\xvec))
 \end{equation}
 where we have used the convention that $\vec{\beta}\in\R^{\pardim{f^a_{p(e)}}\times \abs{V^b}}$ can be expressed as $\vec{\beta} = (\vec{\beta_{v^b_1}}, \ldots, \vec{\beta_{v^b_{\abs{V^b}}}})$ and for every $i\in\Nset{\abs{V^b}}$, $\vec{\beta_{v^b_i}}\in \R^{\pardim{f^a_{p(e)}}}$. We define $F^{2,c} = \bigcup\limits_{e\in E^c}f^2_e$. A symmetric definition can be given for $e\in E^d$ yielding $F^{2,d} = \bigcup\limits_{e\in E^d}f^2_e$. We define $F^2 = F^{2,c}\cup F^{2,d}$ and we say $D^2=(V, E, F^2)$. We define
 \begin{equation}
    D^a\Box D^b = D^2.
 \end{equation} We call $\Box$ \textbf{the modified product}.
\end{definition}

\FloatBarrier

Given two models, their naive and modified products will have identical vertex and edge sets. In fact, the only difference between the two is their flow functions. 

\subsection{Example Naive and Modified Products}

We now describe specific examples of naive and modified products. We will note here that in our example models we have erred on the side of simplicity. We occasionally omit minor details which, while important for the actual job of modeling, are unnecessary and potentially distracting for the purpose of clearly communicating our point.  

\begin{figure}
    \centering
    \includegraphics[width=\textwidth]{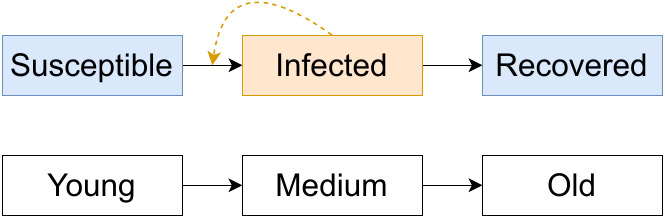}
    \caption{A standard SIR model and a simple age stratification model. Orange shading denotes the infectious compartment, while blue shading denotes non-infectious compartments.}
    \label{fig:sir_age_strat}
\end{figure}

\begin{figure}
    \centering
    \includegraphics[width=\textwidth]{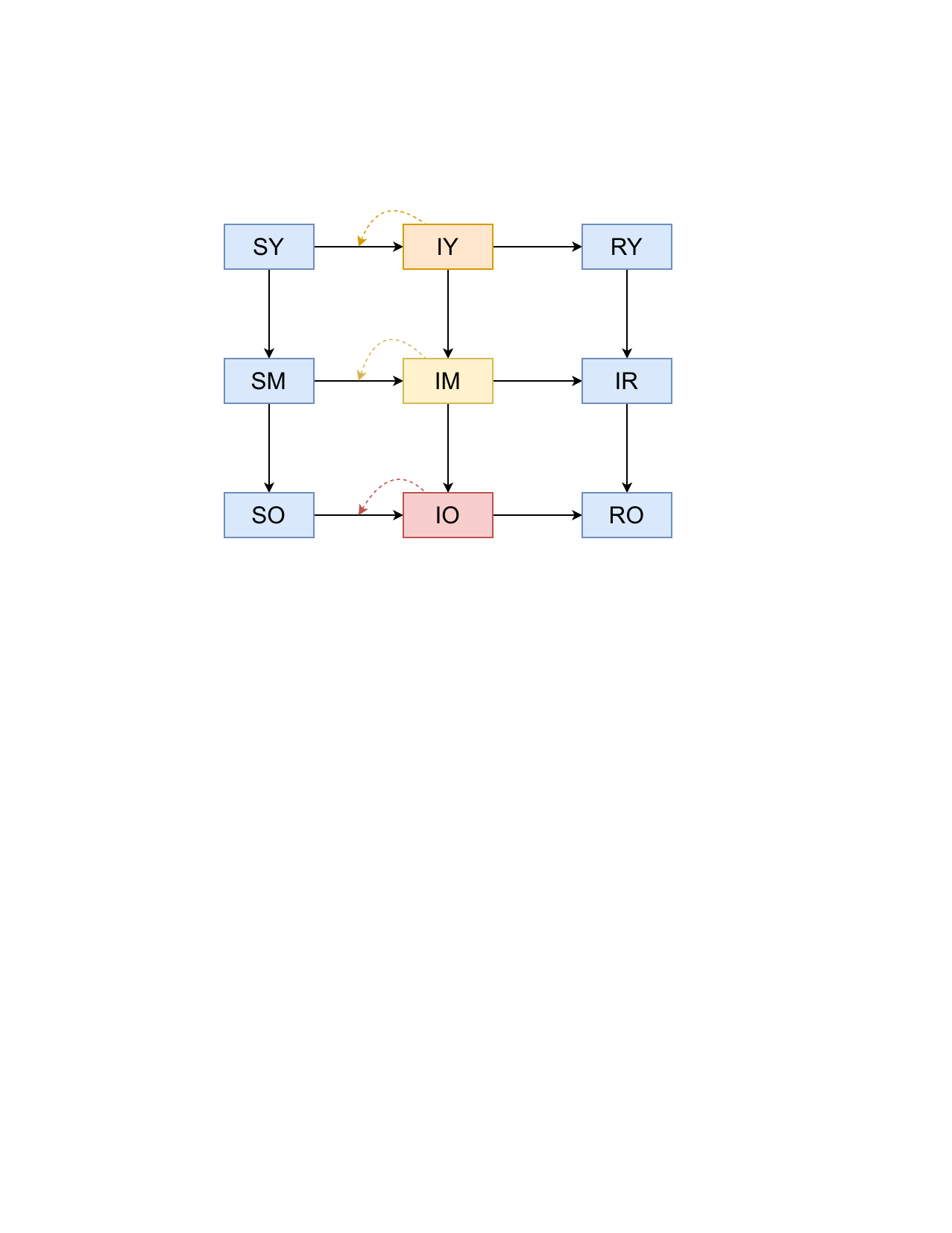}
    \caption{The naive product of the two models from Figure \ref{fig:sir_age_strat}. Blue denotes non-infectious compartments, yellow/orange/red denote infectious compartments. The force of infection is only influenced by the infected population within the same age stratum. In this example, people of different age groups have no contact (or very limited contact) with each other.}
    \label{fig:naive_product}
\end{figure}

\begin{figure}
    \centering
    \includegraphics[width=\textwidth]{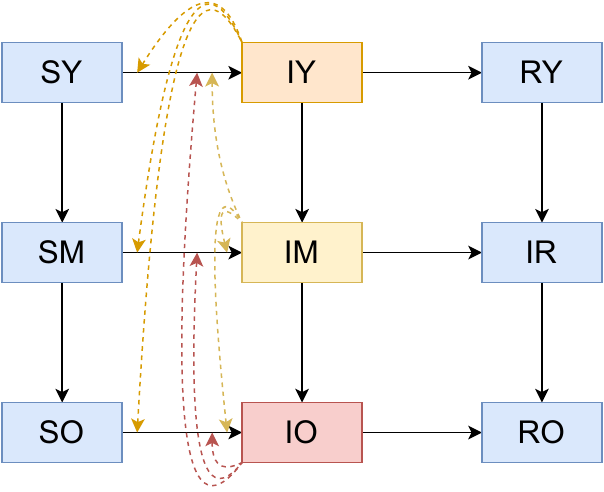}
    \caption{The modified product of the two models from Figure \ref{fig:sir_age_strat}. Unlike in Figure \ref{fig:naive_product}, individuals make epidemiological contacts across age strata, so the force of infection for each age stratum is influenced by the infected population in all age strata.}
    \label{fig:modified_product}
\end{figure}

To illustrate the difference between naive and modified products, consider the two compartmental models in Figure \ref{fig:sir_age_strat}. The top model tracks the epidemiological status of a population of susceptible, infected, and recovered individuals, and can be written in our notation as $G = (V^g, E^g, F^g)$ where $V^g = \{ S, I, R\}$, $E^g=\{(S, I), (I, R)\}$, and $F^g = \{S\frac{\beta I}{S+I+R}, I\gamma \}$. The bottom model divides the population into three age groups, young, medium, and old, and can be written as $H = (V^h, E^h, F^h)$ where $V^h = \{Y, M, O\}$, $E^h = \{(Y, M), (M, O)\}$, and $F^h = \{\frac{Y}{\mu}, \frac{M}{\mu}\}$, where $\mu$ is the expected residency time for people in the young and medium age groups. We can then compute both $G\Box H = M$ (Figure \ref{fig:modified_product}) and $G\boxminus H = N$ (Figure \ref{fig:naive_product}). Note that $V^m = V^n$ and $E^m = E^n$; the only difference between $M$ and $N$ is the difference between $F^m$ and $F^n$. So
    \begin{equation}
        V^m = V^n = \{SY, SM, SO, IY, IM, IO, RY, RM, RO\}
    \end{equation}
    \begin{multline}
        E^m = E^n = \left\{ (SY, IY), (IY, RY), (SM, IM), (IM, RM), (SO, IO), (IO, RO), \right.\\ \left. (SY, SM), (SM, SO), (IY, IM), (IM, IO), (RY, RM), (RM, RO)\right\}
    \end{multline}
    \begin{multline}\label{normex1}
        F^m = \left\{SY\left(\frac{\beta_1 IY}{SY+IY+RY} + \frac{\beta_2 IM}{SM+IM+RM}+\frac{\beta_3 IO}{SO+IO+RO}\right), IY\gamma_1,\right.\\ \left.SM\left(\frac{\beta_1 IY}{SY+IY+RY} + \frac{\beta_2 IM}{SM+IM+RM}+\frac{\beta_3 IO}{SO+IO+RO}\right) ,IM\gamma_2,\right. \\ \left. SO\left(\frac{\beta_1 IY}{SY+IY+RY} + \frac{\beta_2 IM}{SM+IM+RM}+\frac{\beta_3 IO}{SO+IO+RO}\right), IO\gamma_3,\right. \\ \left.  \frac{SY}{\mu}, \frac{SM}{\mu}, \frac{IY}{\mu}, \frac{IM}{\mu}, \frac{RY}{\mu}, \frac{RM}{\mu} \right\}
    \end{multline}
    \begin{multline}\label{normex2}
        F^n = \left\{SY\frac{\beta_1 IY}{SY+IY+RY}, IY\gamma_1,  SM\frac{\beta_2 IM}{SM+IM+RM}, IM\gamma_2, SO\frac{\beta_3 IO}{SO+IO+RO}, IO\gamma_3,\right. \\ \left. \frac{SY}{\mu}, \frac{SM}{\mu}, \frac{IY}{\mu}, \frac{IM}{\mu}, \frac{RY}{\mu}, \frac{RM}{\mu} \right\}
    \end{multline}

\FloatBarrier

\FloatBarrier
The term “naive product" is not pejorative; in this specific example the modified product is likely to be preferred because people of all ages commonly interact with each other. In some scenarios, however, the naive product would be preferred. In the case of spatial stratification, for example, one might want to use the naive, the modified, or another alternative depending on the specifics of the epidemiological system. At first glance the naive product seems promising because it incorporates the idea that people at different physical locations cannot interact and so do not infect one another. This, however, assumes that the model in question simulates movement explicitly which is not always the case. In product models that do simulate movement explicitly \cite[e.g.][]{mohammadi2023importation}, the flows between different locations are included in the original factor model describing spatial structure. Other models, such as \cite{dietz1995structured}, model movement implicitly. In this sort of model a person's location stratum might determine where they live but the possibility of their movement to another location is represented as a contact rate between people in their home stratum and the stratum they might visit. In this case the modified product would seem the most appropriate. However, the situation can be even more complex; for example, depending on the scales involved, it may be beneficial to allow contact between people who are either in the same location \emph{or} in neighbouring locations, for some appropriate definition of a neighbourhood. In this case we want to go beyond the naive or modified products to some kind of \define{generalized product}.

\section{Generalized Product}\label{general}
The naive product restricts people in each stratum to interacting only with other people in the same stratum; the modified product allows people in any strata to interact. We propose a new product that allows for people in each stratum to interact with people in an arbitrary subset of the other strata. This allows the creation of a model where people at a given location can interact at the same location or neighboring locations. The following definition is very similar to Definition \ref{products}; however unlike Equation \ref{modified_eq}, where the sum is over all vertices in a factor model, the sum in this definition will only include a subset of a factor model's vertices.

\begin{definition}[Generalized Product]
    Let $D^a = (V^a, E^a, F^a)$ and $D^b = (V^b, E^b, F^b)$ be two compartmental models and let $V$, $E$, $E^c$, and $E^d$ be as in Definition \ref{cpd}. As in Definition \ref{compmodel} we know that every edge $e\in E$ is associated with an edge in one factor model and a vertex in the other; we use $p(e)$ to denote the associated factor model edge and $s(e)$ to denote the associated factor model vertex. For every $e\in E^a$ there are functions $f^a_{e}\in F^a$ and $f_{r,e}\in \fr(\abs{V^a})$, as well as a filtering function $\tauf_{e}\in\Upsilon(\abs{V^a}, 1)$ such that
    \begin{equation}
        f^a_{e}(\betavec, \xvec) = \tauf_e(\xvec)\cdot f_{r,e}(\betavec, \xvec).
    \end{equation}
        For every $v^b\in V^b$ let $\taua_{v^b}$ be the filtering function from $V$ to $V^a\times v^b$ and let $V_{v_b}\subset V^b$ denote those vertices in $V^b$ that belong to strata that interact with the $V^a\times v^b$ stratum. For every $e\in E^c$ and $\vec{x}\in\Rnn^{\abs{V}}$ we define $f_e:\R^{\pardim{f^a_{p(e)}}\times \abs{V_{v_b}}}\times\Rnn^{\abs{V}}\rightarrow\Rnn$ by
     \begin{equation}
        f_e(\vec{\beta}, \vec{x}) = \tauf_{p(e)}(\taua_{s(e)}(\xvec))\sum\limits_{v^b\in V_{v_b}} f_{r,p(e)}(\betavec_{v^b},\taua_{v^b}(\xvec))
     \end{equation}
     We define $F^{c} = \bigcup\limits_{e\in E^c}f_e$. A symmetric definition can be given for $e\in E^d$ yielding $F^{d} = \bigcup\limits_{e\in E^d}f_e$. We define $F = F^{c}\cup F^{d}$ and we say $D=(V, E, F)$. We define
    \begin{equation}
        D^a\boxdot D^b = D.
    \end{equation} We call $\boxdot$ \textbf{the generalized product}.
\end{definition}

\subsection{Example Generalized Products}

\begin{figure}
    \centering
    \includegraphics[width=\textwidth]{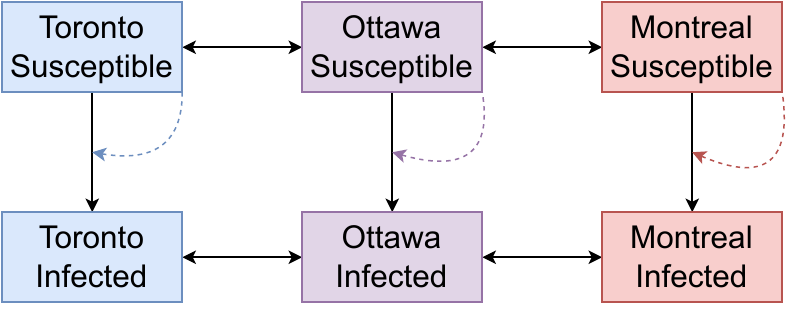}
    \caption{The naive product of an SI model with location model including Toronto, Ottawa, and Montreal. Notice the force of infection at any given location is influenced is determined by the size of the infectious population at the same location. This approach is useful when movement between locations is being modeled explicitly}
    \label{fig:spat_n}
\end{figure}

\begin{figure}
    \centering
    \includegraphics[width=\textwidth]{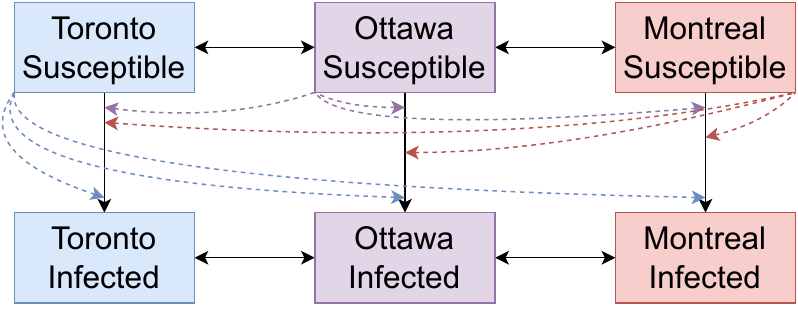}
    \caption{The modified product of an SI model with location model including Toronto, Ottawa, and Montreal. Notice that the size infectious population at each location influences the force of infection at every location. This approach is useful when movement between location is modeled implicitly, for example through non-zero contact rates between populations at different locations}
    \label{fig:spat_m}
\end{figure}

\begin{figure}
    \centering
    \includegraphics[width=\textwidth]{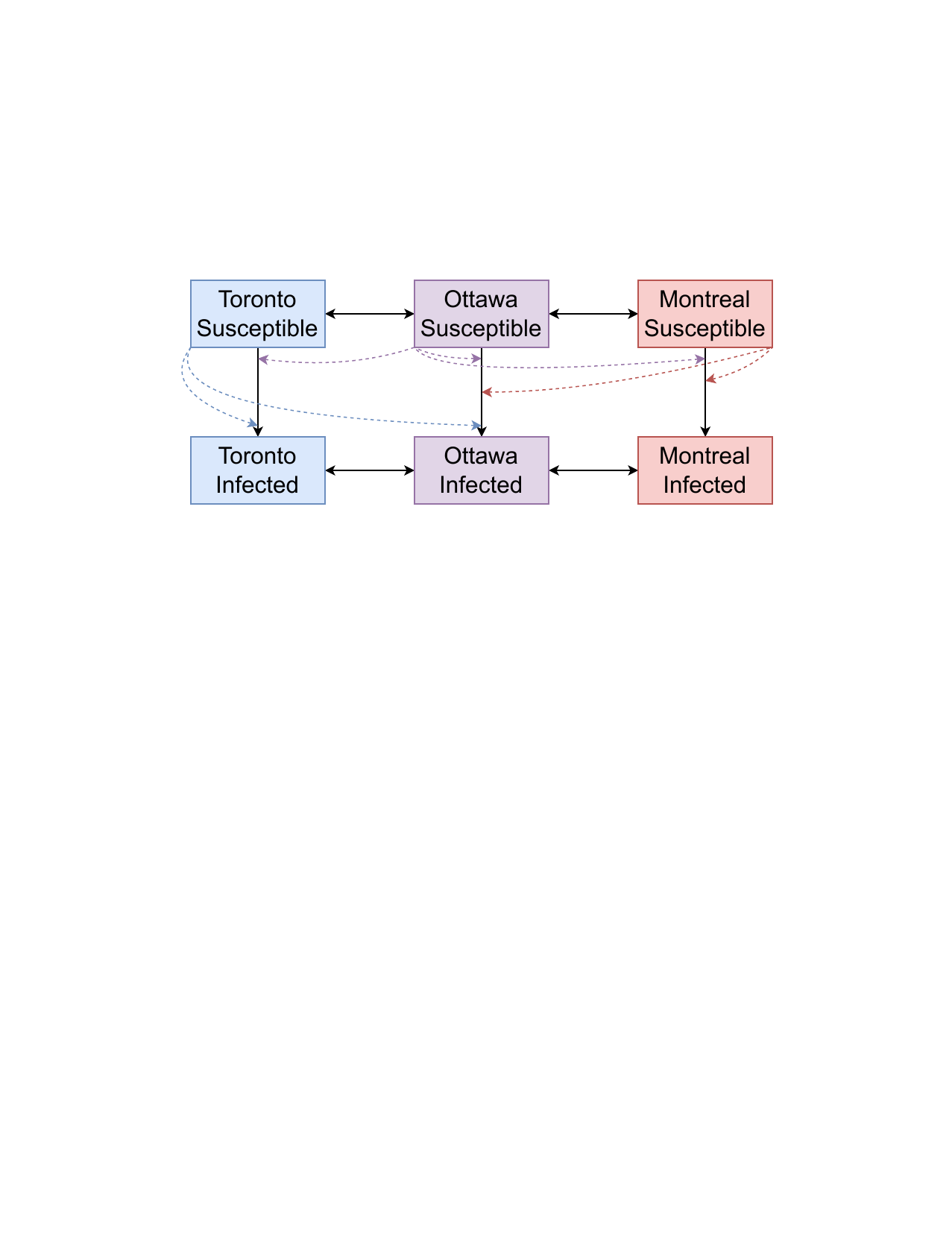}
    \caption{A generalized product of an SI model with location model including Toronto, Ottawa, and Montreal. Notice that the force of infection at a given location depends on the size of the infectious population at the same location \textit{and} at neighboring locations. This approach has a wide variety of applications, for example when there is significant variance in the distance between different locations}
    \label{fig:spat_g}
\end{figure}

Below we show three different ways an SI model could be stratified with location. Figures \ref{fig:spat_n} and \ref{fig:spat_m} show the naive and modified products respectively. Figure \ref{fig:spat_g} shows one example of a generalized product where interactions can only occur within a single geographic region or between neighboring regions. So for example an infected person in the Toronto region could infect a susceptible person in Toronto or Ottawa but not one in Montreal. However, an infected person in Ottawa could infect a susceptible person anywhere. 

\FloatBarrier

\FloatBarrier

\section{Uncooperative Examples}\label{unco}
While the operations defined above allow us to construct a wide range of compartmental models by taking products of simpler factor models, they cannot account for every possible model. In this section we discuss a number of examples where products alone, as they appear in this article, are insufficient.

\subsection{Models with alternate functional forms}\label{aff}
Although we have tried in this article to make the functional forms we use as general as possible there are still certain approaches that do not fit neatly into the formalism we have developed. Fortunately the difficulties that arise from this can typically be overcome by designing a new model product that possesses the desired properties. As we stated in Section \ref{intro} in this paper we have followed the approach of Worden and Porco when computing the magnitude of flows in a product model. That is, we individually compute the contribution from every stratum and compute the sum. In an alternate approach, we instead take a weighted average of the compartment populations in each stratum and use this new average as the input to the flow rate function. This approach is particularly useful when incorporating inhibitory influences in a model. For example during an epidemic will be more careful if they know hospitals are at capacity than they would be when there are ample medical resources available. We introduce the following product definition:

\begin{definition}[Weighted States Product]
    Let $D^a = (V^a, E^a, F^a)$ and $D^b = (V^b, E^b, F^b)$ be two compartmental models and let $V$, $E$, $E^c$, and $E^d$ be as in Definition \ref{cpd}. As in Definition \ref{compmodel} we know that every edge $e\in E$ is associated with an edge in one factor model and a vertex in the other; we use $p(e)$ to denote the associated factor model edge and $s(e)$ to denote the associated factor model vertex. For every $e\in E^a$ there are functions $f^a_{e}\in F^a$ and $f_{r,e}\in \fr(\abs{V^a})$, as well as a filtering function $\tauf_{e}\in\Upsilon(\abs{V^a}, 1)$ such that
    \begin{equation}
        f^a_{e}(\betavec, \xvec) = \tauf_e(\xvec)\cdot f_{r,e}(\betavec, \xvec).
    \end{equation}
        For every $v^b\in V^b$ let $\taua_{v^b}$ be the filtering function from $V$ to $V^a\times v^b$. For every $e\in E^c$ and $\vec{x}\in\Rnn^{\abs{V}}$ we define $f_e:\R^{\pardim{f^a_{p(e)}}\times \abs{V_{v_b}}}\times\Rnn^{\abs{V}}\rightarrow\Rnn$ by
     \begin{equation}
        f_e(\vec{\beta}, \vec{x}) = \tauf_{p(e)}(\taua_{s(e)}(\xvec)) f_{r,p(e)}\left(\betavec_{s(e)},\sum\limits_{v^b\in V^b} w_{v^b}\taua_{v^b}(\xvec)\right)
     \end{equation}
     where for every $v^b\in V^b$, $w_{v^b}\in\Rnn$ denotes the weight to be given to the $v^b$ strata in the weighted average. We define $F^{c} = \bigcup\limits_{e\in E^c}f_e$. A symmetric definition can be given for $e\in E^d$ yielding $F^{d} = \bigcup\limits_{e\in E^d}f_e$. We define $F = F^{c}\cup F^{d}$ and we say $D=(V, E, F)$. We define
    \begin{equation}
        D^a\triangle D^b = D.
    \end{equation} We call $\triangle$ \textbf{the weighted states product}.
\end{definition}

Alternatively, one might notice that if a flow rate function involves normalization, when computing the contribution to the flow of each stratum individually the normalization is done with respect to the population of the individual strata rather the the total population of the model. One example of this can be seen in Equations \ref{normex1} and \ref{normex2}. To state this difference more clearly we introduce a new function $N$ which simply sums the population of every compartment in a state vector. So for $\xvec\in \Rnn^n$,
\begin{equation}
    N(\xvec) = \sum\limits_{i=0}^{n-1}x_i
\end{equation}
Recalling Definition \ref{flowfunc} many, but not necessarily all, flow functions in a model will have the form
\begin{equation}
    f(\betavec, \xvec) = \tau(\xvec)\cdot f_r(\betavec, \xvec) = \tau(\xvec) \cdot \frac{g(\betavec, \xvec)}{N(\xvec)}
\end{equation}
where $g\in\fr$. If a factor model has flow functions with the above form then, if that model is used in a modified product for example, the form of the related flow function in the product model which is given by Equation \ref{modified_eq} will be
\begin{equation}
        f_e(\vec{\beta}, \vec{x}) = \tauf_{p(e)}(\taua_{s(e)}(\xvec))\sum\limits_{v^b\in V^b} \frac{g(\betavec_{v^b}, \taua_{v^b}(\xvec)}{N(\taua_{v^b}(\xvec))}
\end{equation}
    where in fact, depending on the specific context the desired outcome may be
\begin{equation}
        f_e(\vec{\beta}, \vec{x}) = \frac{\tauf_{p(e)}(\taua_{s(e)}(\xvec))}{N(\xvec)}\sum\limits_{v^b\in V^b} g(\betavec_{v^b},\taua_{v^b}(\xvec))
\end{equation}
This difference suggests yet another type of model product, although a rigorous definition would require altering Definition \ref{flowfunc} to explicitly include a denominator in the form of a flow function. 

Another example where the functional form used in Definition \ref{flowfunc} may require amendment is when investigating models with non-linear incidence rates. In such cases flow functions related to the force of infection in a model may have an additional factor associated with them yielding the form:

\begin{equation}
    f(\betavec, \xvec) = \tauf(\xvec)\cdot f_r(\betavec, \xvec) = \frac{\tauf(\xvec)^\zeta}{N(\xvec)^{\zeta + 1}} \cdot g(\betavec, \xvec)
\end{equation}

Defining a space of flow functions which allows some, but not all, of the flows in a model to employ these variations in form is a non-trivial challenge and a significant reason why it is difficult to build a simulation engine for a product model without the supervision of a trained programmer.

\subsection{Models with Testing}\label{testing}

\begin{figure}
    \centering
    \includegraphics[width=\textwidth]{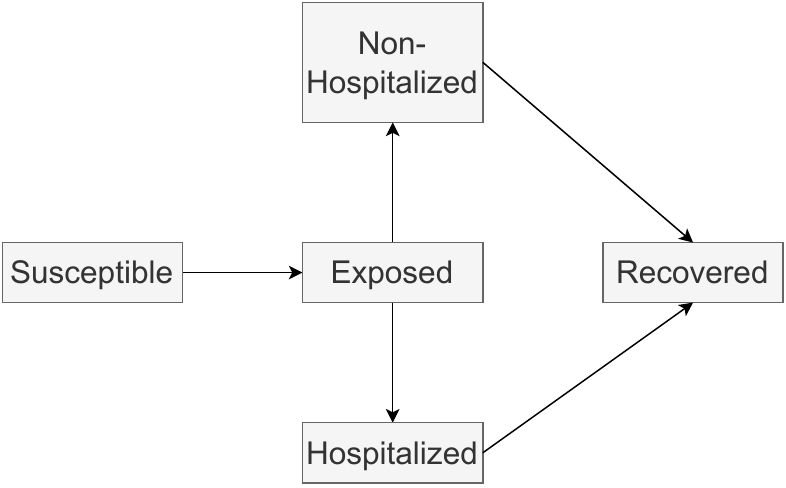}
    \caption{A simple epidemiological model that we will expand to include testing. In this model, some exposed individuals will develop asymptomatic or mild illness, in which case they stay in the community during their infectious period (and potentially transmit to others); those who instead develop severe illness will be hospitalized. (This model allows neither for within-hospital transmission nor for disease-induced mortality either inside or outside the hospital.)}
    \label{fig:testify_epi}
\end{figure}

\begin{figure}
    \centering
    \includegraphics[width=\textwidth]{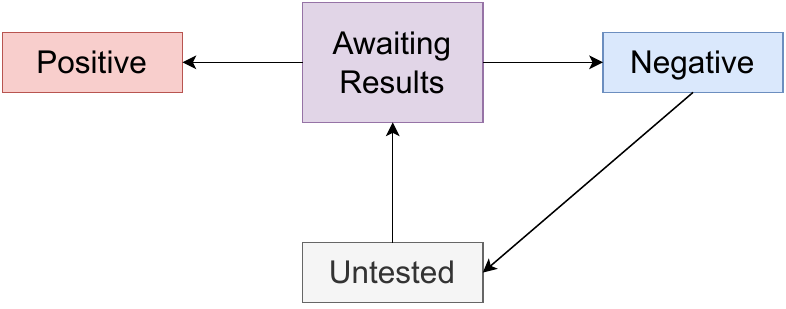}
    \caption{A simple testing model. Individuals who test negative will, over time, revert back to the “untested" status. This is not the case for those that test "positive"; at least during the early stages of the COVID-19 pandemic, someone who had tested positive for COVID-19 would assume that they were immune and would not be re-tested even if they developed COVID-like symptoms.}
    \label{fig:testify_states}
\end{figure}

\begin{figure}
    \centering
    \includegraphics[width=\textwidth]{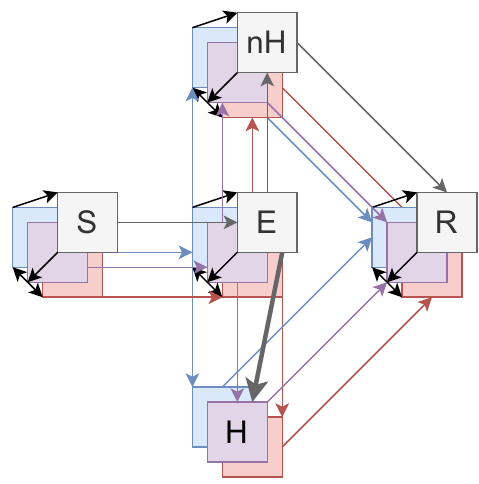}
    \caption{The desired result of combining Figure \ref{fig:testify_epi} with Figure \ref{fig:testify_states}. Note the missing grey ``untested'' box associated with the hospital location; exposed individuals going into the hospital (enlarged, grey downward arrow starting at $E$) flow into the purple ``awaiting results'' subcompartment.}
    \label{fig:testify_desired}
\end{figure}

One such example (where model products alone cannot produce the desired result) involves modeling the effects of testing for infection, inspired by the dynamics of testing during the COVID-19 pandemic. One example of a model that includes the effects of testing can be found in \cite{gharouni2022testing}. Consider the epidemiological model in Figure \ref{fig:testify_epi} and the testing process depicted in Figure \ref{fig:testify_states}. The modified product of these to models includes a compartment for untested individuals at the hospital. However, this product is not what we want (Figure \ref{fig:testify_desired}). 
The key difference is that untested individuals entering the hospital are typically tested (i.e., moved from ``untested'' to ``awaiting results''); our model world assumes that they always are.
Therefore, the “untested hospitalized" compartment in product model is empty  and should be eliminated; the flow that goes to that compartment should instead be directed to the “hospitalized/awaiting test result" compartment. Constructing the desired model would thus require an extra step to remove the superfluous compartment.

\FloatBarrier

\subsection{Multistrain Models and a Weak Product}\label{wp}
Many epidemics involve multiple co-circulating strains of the same pathogen \citep{gog2002dynamics, williams2021localization}. In the case of COVID-19 such variants have significant implications for the efficacy of vaccines\citep{abu2021effectiveness, koyama2020emergence} and diagnostic tests \citep{vasireddy2021review}. In more complex models, including multiple strains rapidly inflates the size of both the state space and the parameter space \citep{kryazhimskiy2007state}. One way to limit the size of these unwieldy models while continuing to include the effects of multiple strains in our model is to disallow the possibility of \emph{superinfection} (i.e. an individual being infected with multiple strains at the same time). It would therefore be useful to define a \emph{weak product} similar to the operations proposed by \cite{worden2017products} but which excludes all states corresponding to a superinfected status. Below we propose such a product which is satisfactory for producing two-strain models but, as we shall see, fails for more than two strains.

\begin{definition}[Weak product]
    Suppose $A$ and $B$ are two compartmental models; let $S_a\subset V(A)$ and $S_b\subset V(B)$ denote the set of all vertices with either no inflows (i.e. \emph{sources}) or no outflows (i.e. \emph{sinks}) in $A$ and $B$ respectively. Let $I_a=V(A)\setminus S_a$ and $I_b=V(B)\setminus S_b$. We define
    \begin{equation}
        V_c = (S_a\times S_b) \cup (S_a\times I_b) \cup (S_b\times I_a).
    \end{equation}
    We define
    \begin{equation}
        E_c^a = \left\{((w,x),(y,x))\in V_c\times V_c\vert (w,y)\in E(A) \right\},
    \end{equation}
    \begin{equation}
        E_c^b = \left\{((w,x),(w,z))\in V_c\times V_c\vert (x,z)\in E(B) \right\},
    \end{equation}
    and
    \begin{equation}
        E_c = E_c^a\cup E_c^b.
    \end{equation}
    For every $e\in E_c^a$, let $p(e) = (w,y)\in E(A)$. For every $e\in E(A)$ there are functions $f^a_{e}\in F(A)$ and $f_{r,e}\in \fr(\abs{V(A)})$, as well as a filtering function $\tauf_{e}\in\Upsilon(\abs{V(A)}, 1)$ such that
    \begin{equation}
        f^a_{e}(\betavec, \xvec) = \tauf_e(\xvec)\cdot f_{r,e}(\betavec, \xvec)
    \end{equation}
    For every, $v_b\in S_b$ let $\taua_{v_b}$ be the filtering function from $V$ to $V(A)\times v_b$. For every $e\in E_c^a$ we define
    \begin{equation}
        f_e(\betavec, \xvec) = \sum\limits_{v_b\in S_b}\tauf_{p(e)}(\taua_{v_b}(\xvec))f^a_{p(e)}(\betavec_{v_b}, \taua_{v_b}(\xvec))
    \end{equation}
    We define $F_c^a = \bigcup\limits_{e\in E_c^a}f_e$. A symmetric definition can be given for $e\in E_c^b$ yielding $F_c^b = \bigcup\limits_{e\in E_c^b}f_e$. We define $F_c = F_c^a\cup F_c^b$ and we say $C = (V_c, E_c, F_c)$. We define
    \begin{equation}
        A\boxplus B = C
    \end{equation} We call $\boxplus$ \textbf{the weak product}.
    
\end{definition}

\begin{figure}
    \centering
    \includegraphics[width=\textwidth]{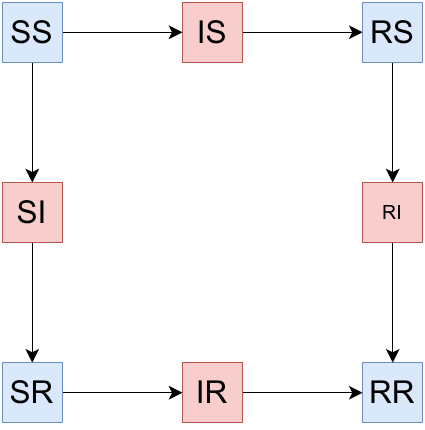}
    \caption{A two-strain SIR model admitting no superinfection. Red Compartments indicate an infectious population whereas the population in blue compartments are not infectious}
    \label{fig:ms_ns_2d}
\end{figure}

\begin{figure}
    \centering
    \includegraphics[width=\textwidth]{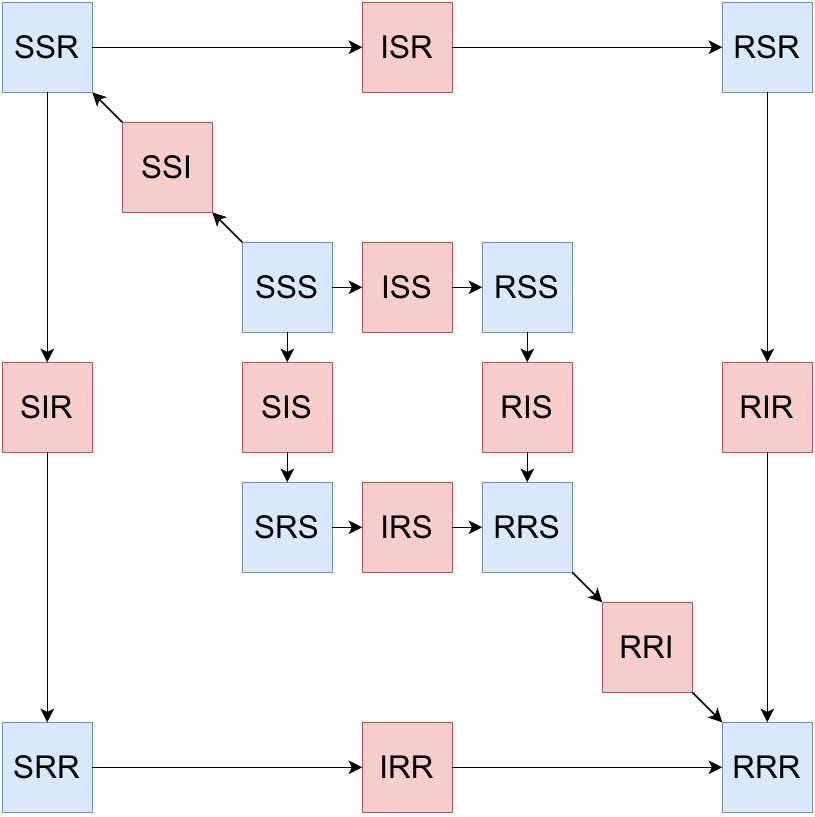}
    \caption{A model corresponding to the product $(A\boxplus B)\boxplus C$. Notice that the “SRS" and “RSS" compartments are not sinks or sources in $A\boxplus B$ hence why they have no paths to “SRR" and “RSR" respectively. }
    \label{fig:wp_act}
\end{figure}

\begin{figure}
    \centering
    \includegraphics[width=\textwidth]{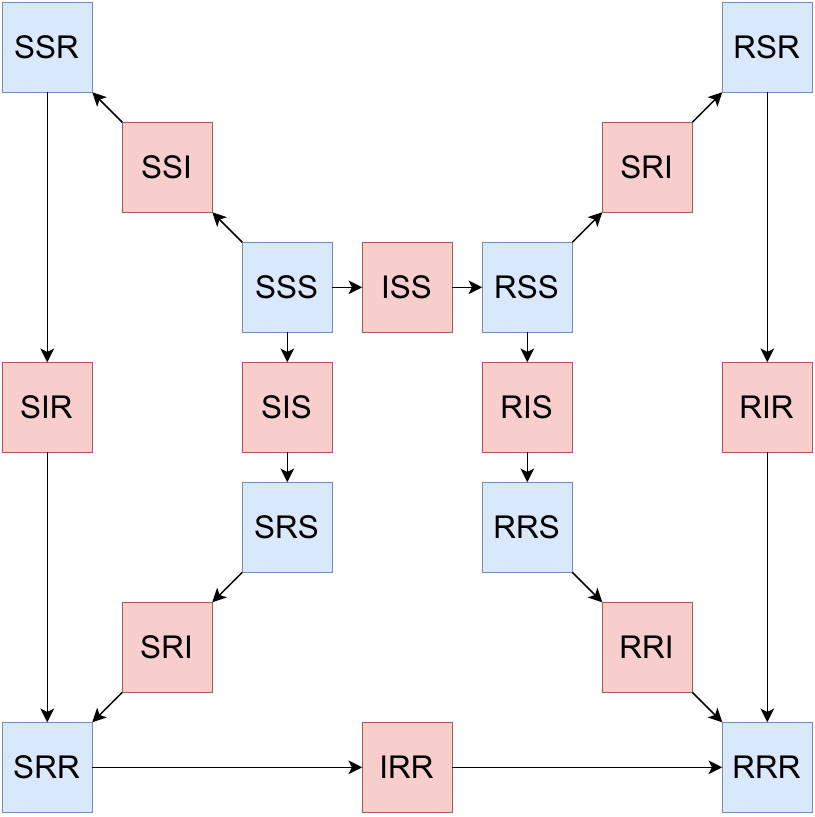}
    \caption{A model corresponding to the product $A\boxplus (B\boxplus C)$. Notice that “SRS" and “SSR" compartments are not sources or sinks in $(B\boxplus C)$ hence why they have no paths to “RRS" and “RSR" respectively}
    \label{fig:wp_nonasoc}
\end{figure}

\begin{figure}
    \centering
    \includegraphics[width=\textwidth]{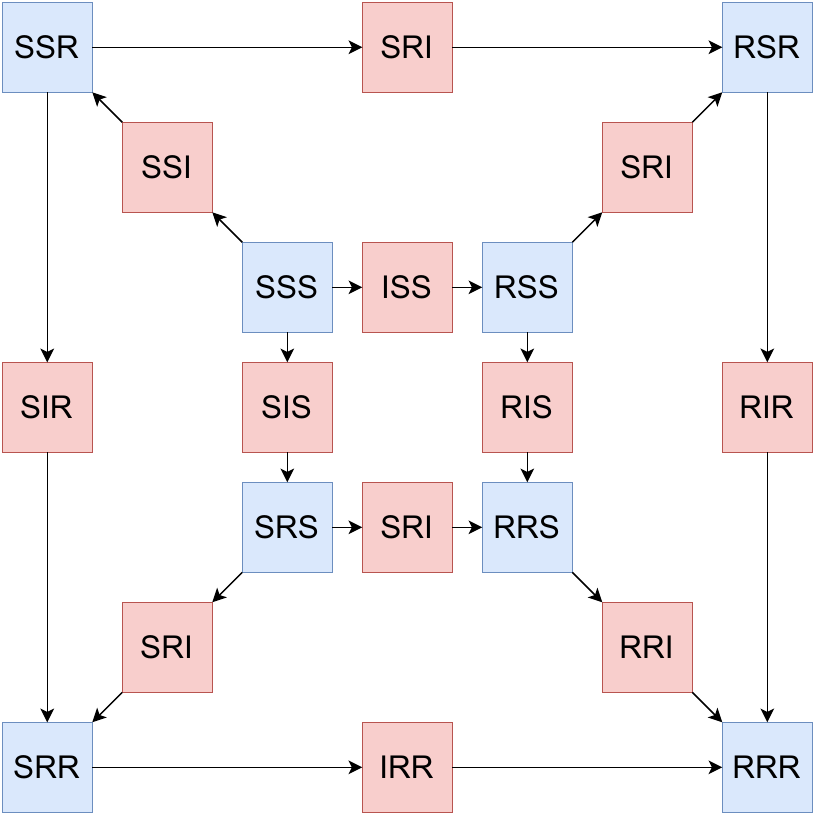}
    \caption{A three strain SIR model admitting no superinfection. This model cannot be constructed using only the products defined in this article}
    \label{fig:ms_ns_3d}
\end{figure}

\begin{figure}
    \centering
    \includegraphics[width=\textwidth]{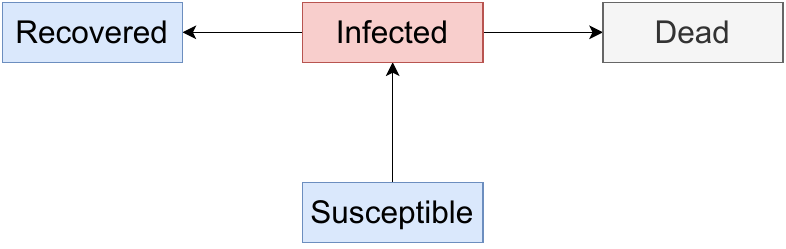}
    \caption{A single strain model with two sinks and one source}
    \label{fig:msms}
\end{figure}

\begin{figure}
    \centering
    \includegraphics[width=\textwidth]{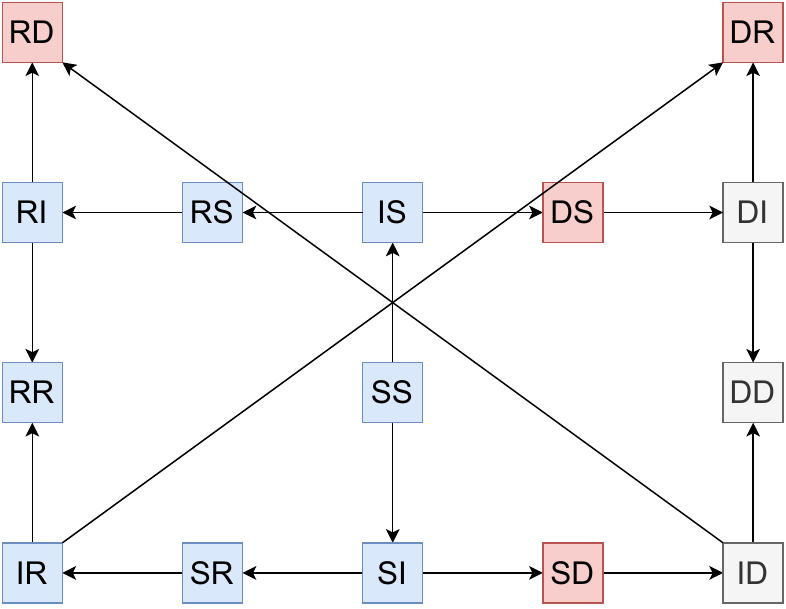}
    \caption{The weak product of two of the single strain models depicted in Figure \ref{fig:msms}. Notice that the grey compartments are superfluous as they correspond to changes in infection status occurring after death}
    \label{fig:msms2}
\end{figure}

Figure~\ref{fig:ms_ns_2d} depicts a two-strain SIR model without super-infection; this corresponds to the weak product of two SIR models. Figures \ref{fig:wp_act} and \ref{fig:wp_nonasoc} depict two different results for the weak product of three SIR models; note that the difference between them results from changing the order in which products are done. If, for example, we denote the models of the red, yellow, and purple strains by $A$, $B$, and $C$ respectively then Figure \ref{fig:wp_act} depicts $(A\boxplus B)\boxplus C$ and Figure \ref{fig:wp_nonasoc} depicts $A\boxplus (B\boxplus C)$. Figure \ref{fig:ms_ns_3d} depicts the desired result for a three-strain SIR model with no super-infection. It is possible to create a version of the weak product defined above that will produce the model shown in Figure \ref{fig:ms_ns_3d}; however it requires us to distinguish between compartments that are global sources or sinks and compartments that are sources or sinks with respect to one of the three strains specifically. That is to say, while a global sink must have no outflows, a weaker condition says that a compartment is a sink with respect to a specific pathogen if every compartment that can be reached via the outflow has the same infection status with respect to that pathogen as the original compartment. Programmatically we achieve this by introducing a concept of `labeled partitions' which separates the vertices of the model into disjoint sets corresponding to the vertices' status with respect to a specific pathogen. Each dimension of stratification in the model corresponds to a different labeled partition with each stratum corresponding to a different disjoint set. In this way we can define sources and sinks with respect to a specific set of labels rather than globally. For example, we can say a compartment $A$ is a sink with respect to a specific labelled partition if every compartment that can be reached after being in $A$ is in the same set as $A$. Figure \ref{fig:msms} outlines a compartmental model with one source compartment but two sink compartments and Figure \ref{fig:msms2} shows the weak product of two such models.
An unfortunate aspect of this product model is that several of the compartments can only be reached by individuals after they are already dead (!).
If there are relatively few such compartments a modeler may choose simply to leave them in the model and treat them all as a single compartment. But if there are many such “zombie compartments", or if computational efficiency is a pressing concern, they could be removed  from the model.

\FloatBarrier

\FloatBarrier
\section{Conclusion}\label{conc}

Adding new strata to simple epidemiological models is closely related to taking the Cartesian product of digraphs. Modellers who want to combine sets of simple models into a single large stratified model would benefit from a toolkit based on well-defined mathematical operations. This toolkit must contain a variety of operations representing a useful subset of the numerous ways that separate strata in a model can interact. We have developed a mathematical formalism for defining such operations and used it to restate two previously proposed model operations, the naive and modified products, which represent extremes of a spectrum of interactions between strata. The naive product corresponds to the case where different strata never interact, while the modified product corresponds to scenarios where any stratum can interact with any other stratum. We generalize these previously proposed operations to a third operation that allows any level of interaction between model strata, for example to construct geographically stratified models where interactions can occur within a single location and its neighbours but not more distantly.

 Modelers employ a wide variety of functional forms; it is not always clear how to adjust them to accommodate new levels of stratification. For example, the flow functions in a factor model may involve normalization by the total population size. When generalizing to a product model, modelers need to decide whether the correct denominator for any given term is the total population of the model or the population of an individual stratum (either option might be correct depending on the situation). It seems likely that these issues could be resolved by modifying Definition \ref{flowfunc}. For example, one could explicitly include numerator and denominator terms, or even an arbitrary number of terms each potentially drawing its arguments from a different subset of the state space. We have refrained from exploring such possibilities in this paper.

Several challenges remain for anyone wishing to further develop a model construction toolkit. We have paid little attention to parameter space and the question of how to use knowledge about factor model parameters to draw conclusions about product model parameters. It would be convenient to have a catalog of the most common ways of generalizing factor model parameters to product models so that modelers aren't required to reinvent the procedure every time. Many models (e.g. models with infection status testing) also have asymmetries in their structure that cannot be reproduced with Cartesian-like products. This suggests the need for addition-like operations to supplement the multiplication-like operations discussed in this paper. In fact such operations already exist in the category theoretic approach to model operations, which is one reason why it could be a worthwhile project to unite the category theory and graph theory approaches. In our understanding this would involve finding so-called ``type-graphs'' that cause the category theoretic operations known as ``pull-backs'' and ``push-outs'' to  reproduce the results of graph theory operations.

We are heavily motivated by the desire to develop software to facilitate model construction. One insight of our investigations is the utility of a system of so-called ``labeled partitions'', which divide the compartments of a model into mutually exclusive groups. Each group in such a division will contain all compartments that are in the same level of some dimension of stratification and the groups can be labelled accordingly. By applying several such divisions to a model, one for each dimension of stratification, it becomes possible to specify important subsets of the model compartments. Using this system of labels and partitions provides an easy way to address issues like the non-commutativity of the weak product and the presence of ``zombie compartments'' discussed in Section \ref{wp}.

Although theoretical and practical challenges with the application of binary operations on model space remain, our approach forms the basis of a powerful toolkit for the construction of compartmental models.

\section{Acknowledgements}

This project was supported by the Canadian Network for Modelling Infectious Diseases (\href{https://canmod.net/}{CANMOD}), which is funded through the Emerging Infectious Disease Modelling programme of the Natural Sciences and Engineering Research Council of Canada (NSERC).

\section{Declarations}

The authors declare that they have no competing interests.


\end{document}